\documentclass[pra,aps,twocolumn,superscriptaddress]{revtex4}
\usepackage{graphicx}
\usepackage{amsmath}
\usepackage{amssymb}

\newcommand{\ket}[1]{|#1\rangle}
\newcommand{\bra}[1]{\langle#1|}

\newcommand{\eq}{\begin{equation}}
\newcommand{\fine}{\end{equation}}

\begin{document}

\title{Measurement-induced quantum operations on multiphoton states}

\author{Chiara Vitelli}
\affiliation{Dipartimento di Fisica, Sapienza Universit\`{a} di Roma, piazzale Aldo Moro 5, I-00185 Roma, Italy}
\affiliation{Consorzio Nazionale Interuniversitario per le Scienze Fisiche della Materia, piazzale Aldo Moro 5, I-00185 Roma, Italy}
\author{Nicol\`{o} Spagnolo}
\affiliation{Dipartimento di Fisica, Sapienza Universit\`{a} di Roma, piazzale Aldo Moro 5, I-00185 Roma, Italy}
\affiliation{Consorzio Nazionale Interuniversitario per le Scienze Fisiche della Materia, piazzale Aldo Moro 5, I-00185 Roma, Italy}
\author{Fabio Sciarrino}
\email{fabio.sciarrino@uniroma1.it}
\homepage{http://quantumoptics.phys.uniroma1.it}
\affiliation{Dipartimento di Fisica, Sapienza Universit\`{a} di Roma, piazzale Aldo Moro 5, I-00185 Roma, Italy}
\affiliation{Istituto Nazionale di Ottica, Consiglio Nazionale delle Ricerche (INO-CNR), largo Fermi 6, I-50125 Firenze, Italy}
\author{Francesco De Martini}
\affiliation{Dipartimento di Fisica, Sapienza Universit\`{a} di Roma, piazzale Aldo Moro 5, I-00185 Roma, Italy}
\affiliation{Accademia Nazionale dei Lincei, via della Lungara 10, I-00165 Roma, Italy}

\begin{abstract}
We investigate how multiphoton quantum states
obtained through optical parametric amplification can be
manipulated by performing a measurement on a small portion of the
output light field. We study in detail how the macroqubit
features are modified by varying the amount of extracted
information and the strategy adopted at the final measurement
stage. At last the obtained results are employed to investigate
the possibility of performing a microscopic-macroscopic non-locality test free
from auxiliary assumptions. 
\end{abstract}

\maketitle

\section{Introduction}

The possibility of performing quantum operations in order to
tailor quantum states of light on demand has been widely
investigated in the last few years. Several fields of research
have been found to benefit from the capability of generating
fields possessing the desired quantum properties. Non-classical
states of light, such as sub-Poissonian light \cite{Fiur01},
squeezed light \cite{Heer06,Gloc06} or the quantum superposition
of coherent states \cite{Ourj07a,Ourj09}, have been generated in a
conditional fashion. In this context, continuous-variable (CV) quantum information
represents one of the most promising fields where conditional and
measurement-induced non-Gaussian operations can find application.
To this end, quantum interactions can be induced by exploiting linear optics,
detection processes and ancillary states \cite{Fili05}. For
example, the process of coherent photon-subtraction has been
exploited to increase the entanglement present in Gaussian states
\cite{Kita06,Ourj07} and to engineer quantum operations in travelling 
light beams \cite{Fiur09}. Finally, very recently conditional
operations lead to the realization of different schemes for the
implementation of the probabilistic noiseless amplifier
\cite{Ferr10,Xian10,Zava10}, which can find interesting
application within the context of quantum phase estimation
\cite{Usug10}.

Strictly related to the engineering of quantum states of light for applications to quantum
information, there is the problem of beating the decoherence due to
losses which affect quantum resources interacting with an external environment. In the last
few years a large investigation effort has been devoted to the
decoherence process and the robustness of increasing size quantum
fields, realized by non linear optical methods
\cite{DeMa05,Naga07,DeMa08,DeMa09}. Recently quantum phenomena
generated in the microscopic world and then transferred to the
macroscopic one via parametric amplification have been
experimentally investigated. In Ref.\cite{DeMa08} it has been
reported the realization of a resilient to decoherence multiphoton
quantum superposition (MQS) \cite{DeMa09a} involving a large
number of photons and obtained by parametric amplification of
a single photon belonging to a microscopic entangled pair:
$\ket{\psi^{-}}=\frac{1}{\sqrt{2}}\left(\ket{H}_{A}\ket{V}_{B}
-\ket{V}_{A}\ket{H}_{B}\right)$,
where $A,B$ refer to spatial modes $\mathbf{k_{A},k_{B}}$ and the
kets refer to single photon polarization states $\vec{\pi}$
($\pi=H,V$). This process has been realized through a non linear
crystal pumped by a UV high power beam acting as a parametric
amplifier on the single entangled injected photon, i.e. the qubit
$ \left| \phi \right\rangle_{B} $ on spatial mode
$\mathbf{k_{B}}$. In virtue of the unitarity of the optical
parametric amplifier (OPA), the generated state was found to keep
the same superposition character and the interfering properties of
the injected qubit \cite{DeMa98,DeMa05,Naga07} and, by exploiting the
amplification process, the single photon qubit has been converted
into a macro-qubit involving a large number of photons.

\begin{figure}[ht!]
\includegraphics[width=0.5\textwidth]{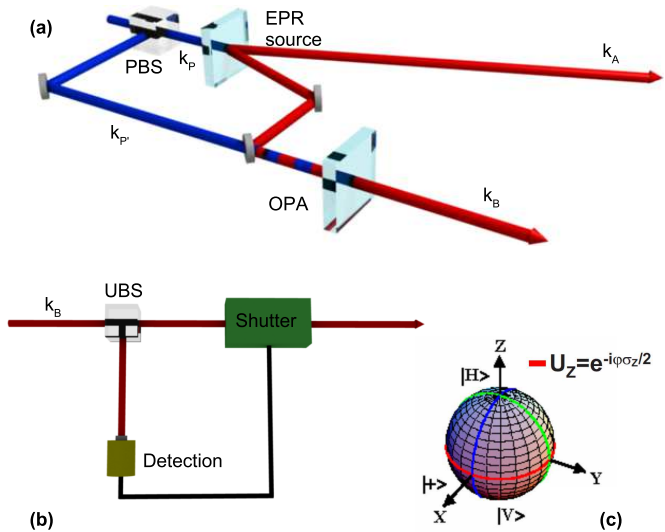}
 \caption{(Color online) (a) Quantum injected optical parametric amplifier 
 (QIOPA) scheme: an entangled photon pair is generated through spontaneous
 parametric down conversion (SPDC). One of the two photon is amplified by
 a non linear crystal, realizing the optimal phase covariant cloning
 of the injected qubit. (b) Scheme of the measurement-induced
 quantum operation process. The field is split by an unbalanced beam
 splitter (UBS), and reflected portion is measured to conditionally active
 the optical shutter placed in the path of the transmitted portion
 of the field. (c) Schematic view of the single photon Bloch sphere.
 The QIOPA device performs the optimal phase covariant process, hence 
 amplifying equally all photons belonging to the equator of the sphere (red line),
 with polarization $\vec{\pi}_{\phi} = 2^{-1/2} (\vec{\pi}_{H} + e^{\imath \phi} 
 \vec{\pi}_{V})$.}
\label{fig:QIOPA_scheme}
\end{figure}

In this paper we consider several strategies for the realization
of measurement-induced quantum operations on these multiphoton
states, generated thought the process of optical parametric
amplification. We investigate theoretically how the measurement
strategies, applied on a part of the multiphoton state before the
final identification measurement, affect the distinguishability of
orthogonal macro-qubits. Such measurements based on the
discrimination of multiphoton probability distributions combine
features of both continuous and discrete variables techniques. The
interest in improving the capability of identifying the state
generated by the quantum injected optical parametric amplifier
(QIOPA) system mainly relies in two motivations: the first one
concerns the development of a discrimination method able to
increase the transmission fidelity of the state after the propagation over a
lossy channel, and hence to overcome the imperfections related to
the practical implementation. Such increased discrimination capability
in lossy conditions could find applications within the quantum 
communication context. The second reason concerns the
scenario in which an appropriate pre-selection of the macro-qubits
could be adopted to demonstrate the micro-macro non-locality, free
from the auxiliary assumptions requested if the filtering
procedure was applied at the final measurement stage. 

In previous papers \cite{Naga07,DeMa08} a probabilistic discrimination method,
the orthogonality-filter (OF), was introduced and successfully applied to
an entanglement test in a microscopic-macroscopic bipartite system. The
application of the OF strategy, acting at the measurement stage, is
indeed not suitable for the demonstration of loophole-free
micro-macro non-locality because of the presence of inconclusive
results \cite{Vite10}. These correspond to the selection of
different sub-ensembles of data, depending on the choice of the
measurement basis. In Ref. \cite{Pope95} Popescu showed that
the pre-selection of data before the final measurement could encompass the problem
of a base dependent filtering of the detected state. More specifically, he
showed that performing a sequence of measurements on each of the two parts
of a bipartite state could reveal the presence of \textit{hidden} non-locality
which was not observable with a single measurement. This method corresponds
to a selection of a sub-ensemble of data \textit{independently} from the 
measurement performed in the non-locality test.
This pre-selection scenario was subsequently extended by Pawlowski \textit{et al.} 
in Ref. \cite{Pawl09} in the more general context of quantum communication.

\begin{center}
\begin{figure*}[ht!]
\includegraphics[width=0.8\textwidth]{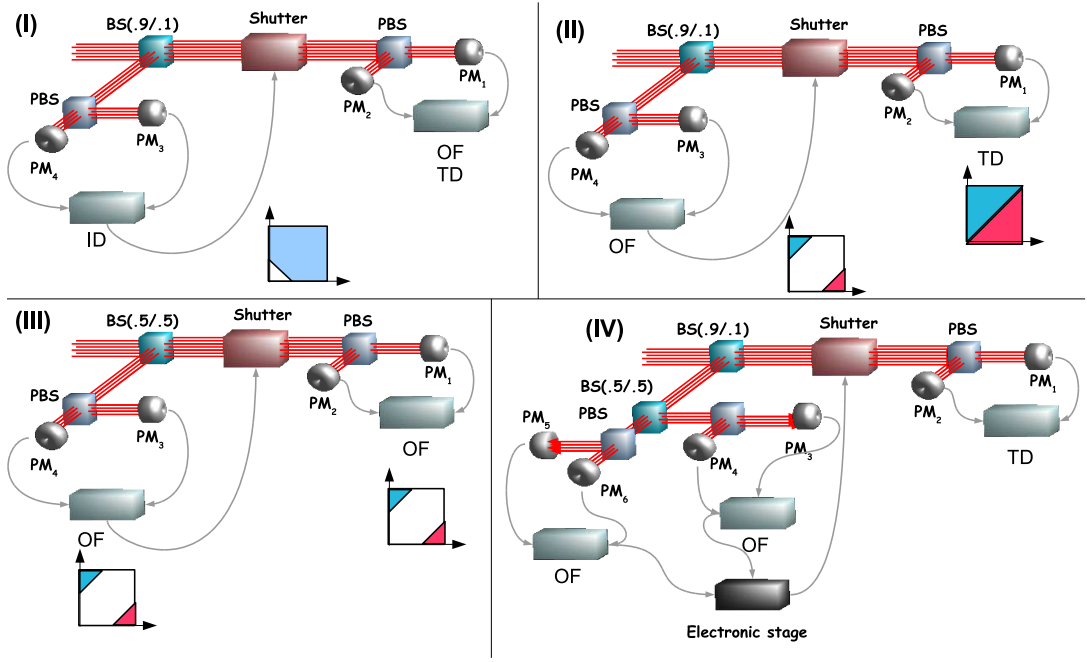}
 \caption{(Color online) Measurement strategies devoted to the manipulation
 of the macro-qubit state: (I) the shutter activation is conditioned to an intensity measurement
 on the reflected portion of the macro state; (II) the small reflected part of the macro state is analyzed in
 polarization and detected through an OF based measurement strategy; (III) the macro state is split in
 two equal parts, and both the reflected and the transmitted components are detected through an OF device;
 (IV) a double basis measurement is performed on the small reflected portion of the macro qubit.}
\label{fig:conceptual_scheme}
\end{figure*}
\end{center}

The capability to generate at the output of the parametric amplifier a quantum state
of large size allows one to act on a small portion of the field in
order to modify the features of the remaining part by a suitable
selection. Conditional manipulation of
quantum states of light depending on measurement carried out on part 
of the beam could increase the capability of discriminating
among the generated multiphoton states, as suggested in \cite{Spag08}. Starting from the original 
proposal of a preselection apparatus in a macro-macro configuration 
of Ref. \cite{DeMa09b}, we consider the particular
case in which a macro-state generated by the QIOPA is split by an
unbalanced beam splitter (UBS) and manipulated by measuring the
state on the reflected mode. The conceptual scheme underlying the
present investigation is shown in figure
\ref{fig:QIOPA_scheme}-(b): a part of the wave-function is
measured and the results of the measurement are exploited to
conditionally activate an optical shutter placed in the
transmitted part. Such shutter, whose realization has been
recently reported in Ref.\cite{Spag08}, is used to allow the
transmission of the optical beam only in presence of a trigger
event, i.e. in this case the results of the measurement performed
in the reflected part of the state. Several detection schemes will
be investigated in this paper. In
fig.\ref{fig:conceptual_scheme}-I is reported the filtering
method based on the intensity pre-selection. As analyzed in more
details in section \ref{sec:distillation}, the signal of the
reflected part of the macro state is analyzed by a threshold
detector. If the measured intensity value is above a certain
threshold, the shutter on the transmitted mode is activated. This
strategy allows to overcome the experimental imperfection related
to the vacuum injection into the amplifier, and hence to
distillate the macro state from the noise belonging to the crystal
spontaneous emission. The application of this scheme has been
already proposed by Stobinska \textit{et al.} in Ref. \cite{Stob09}
to perform a loophole free nonlocality test in a macroscopic-macroscopic
configuration.
In fig\ref{fig:conceptual_scheme}-II is reported the strategy illustrated 
in section \ref{sec:deterministic identification}, based on a probabilistic
discrimination of the reflected macro-qubit part performed by the
OF device. By changing the polarization analysis basis on the
reflected mode, we have investigated how the macro-state
visibility obtained by a dichotomic measurement of the transmitted
state is affected. In fig.\ref{fig:conceptual_scheme}-III is
illustrated the measurement procedure described in section
\ref{sec:probabilistic identification}, in which both the
reflected and the transmitted mode are analyzed by a probabilistic
OF-based measurement. At last, section \ref{sec:non-locality
tests} addresses the case in which the reflected macro qubit part
is measured in two different polarization basis, as shown in
figure \ref{fig:conceptual_scheme}-IV, and the final measurement
is purely dichotomic. This measurement strategy is aimed at the
realization of a non-locality test on the micro-macro photon
state, without any auxiliary assumption. However, we show that
such scheme does not allow to obtain a violation of a Bell's
inequality since the analyzed strategy has the effect of
increasing the correlations present in the micro-macro system only
in a specific polarization basis while suppressing the
correlations in the other basis.

\section{Filtering of the macro-qubit}
\label{sec:distillation}

In this section we discuss the specific preselection scheme, sketched in
Fig. \ref{fig:conceptual_scheme}-(I), proposed in Ref. \cite{Stob09}
in a macro-macro scenario. One of the main experimental
challenge for the realization of the micro-macro system of
Fig.\ref{fig:QIOPA_scheme} is the achievement of spectral, spatial
and temporal matching between the optical mode of the injected
single photon state and the optical mode of the amplifier. In ideal 
conditions, as shown in \cite{Naga07,DeMa08}, the micro macroscopic 
system is realized by the amplification process performed over an 
entangled couple $\ket{\psi^{-}}_{AB}=\frac{1}{\sqrt{2}}\left(
\ket{H}_{A}\ket{V}_{B}- \ket{V}_{A}\ket{H}_{B}\right)$ generated by 
the EPR source in figure \ref{fig:QIOPA_scheme}-(a). In
realistic conditions, the injected micro-micro system is given by:
$\hat{\rho}_{\psi^{-}} = p \vert \psi^{-} \rangle_{AB} \langle
\psi^{-} \vert + (1-p)/2 \hat{I}_{A} \otimes \vert 0 \rangle_{B}
\langle 0 \vert$, where, as said,
$\ket{\psi^{-}}_{AB}$ is the entangled singlet state 
connecting the spatial modes $A$ and $B$, and the
parameter $p$ expresses the amount of mode-matching between the
seed and the amplifier. The micro-macro amplified state $\hat{\rho}_{\Psi^{-}}$ is obtained
after the amplification of the $\hat{\rho}_{\psi^{-}}$ injected state:
$\hat{\rho}_{\Psi^{-}} = (\hat{I}_{A} \otimes \hat{U}_{B}) \hat{\rho}_{\psi^{-}} 
(\hat{I}_{A} \otimes \hat{U}^{\dag}_{B})$, where $\hat{U} = e^{- \imath \hat{H}_{I} t/\hbar}$ is
the time evolution operator associated with the amplifier, defined by the interaction
Hamiltonian $\hat{H}_{I} = \imath \hbar \chi \hat{a}_{H}^{\dag} 
\hat{a}_{V}^{\dag} + \mathrm{H.c.}$. Then in the expression of the number of
photons $N_{\pi_{\pm}}(\varphi)$ generated by the amplifier when
a single photon with equatorial [Fig.\ref{fig:QIOPA_scheme}-(c)] 
polarization state $\vert \phi \rangle$ is injected, the
spontaneous emission has to be taken into account:
$N_{\pi_{\pm}}(\varphi)=p[\bar{m}+\frac{1}{2}(2\bar{m}+1)(1\pm
\cos(\varphi))]+(1-p)\bar{m}$. When the single photon is injected
correctly in the OPA, a pulse with a higher photon number is
generated since stimulated emission processes occur in the
amplifier. The filtering method here presented exploits this
feature to reduce the noise introduced by the spontaneous emission
of the amplifier.

Let us now discuss the propagation of the multiphoton field
produced by the amplifier and the pre-selection procedure obtained
through an intensity threshold detector (ID) and the shutter
device. As shown in fig.\ref{fig:conceptual_scheme}-I, the
amplified state is split by an unbalanced beam splitter (UBS)
$0.90-0.10$ in two parts: the smaller portion on mode
$\mathbf{k_{D}}$ is analyzed by the ID, and the larger one on mode
$\mathbf{k_{C}}$ is conditionally pushed through a polarization
preserving shutter \cite{Spag08}, and measured in polarization by
a dichotomic measurement.
\begin{figure}[ht!]
\includegraphics[width=0.45\textwidth]{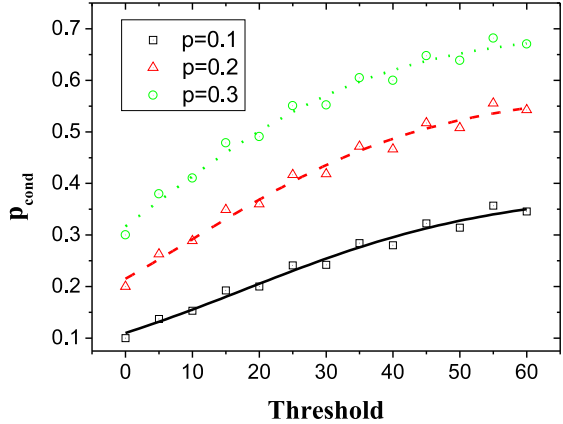}
 \caption{(Color online) Trend of the injection probability as a 
 function of the ID threshold, for different initial values of $p$. The
 nonlinear gain of the amplifier is set at $g=1.5$.}
\label{fig:TD_soglia}
\end{figure}
The ID based filtering strategy allows then to obtain a better
discrimination between the orthogonal macro states, by minimizing
the noise related to the vacuum injection into the amplifier. It is
worth nothing that, at variance with the techniques which will be
introduced in the following sections, the ID action is invariant
for rotation on the Fock space. It indeed selects the same region
of the macro-qubit either in the $\{\vec{\pi}_{+},
\vec{\pi}_{-}\}$ basis either in the $\{\vec{\pi}_{R},
\vec{\pi}_{L}\}$ one. The action of the ID on mode
$\mathbf{k}_{D}$ and of the shutter on mode $\mathbf{k_{C}}$
allows to distill the macro-qubit from the noise generated by the
amplifier and related to the spontaneous emission of the crystal.
In the ideal case, this measurement corresponds to the projection
of the impinging field onto the subspace: $\hat{\Pi}%
_{k}=\sum_{m+n>h}|n\pi ,m\pi _{\bot }\rangle \,\langle n\pi ,m\pi _{\bot }|$, where $%
|n\pi ,m\pi _{\bot }\rangle $ represents a quantum state with $n$
photons with polarization $\pi $ and $m$ photons with polarization
$\pi _{\bot }$. The measurement method is hence based on a
threshold detection scheme, in which the ID clicks only if $n_{\pi
}+m_{\pi _{\bot }}>h$, where $h$ is a threshold conveniently
selected. This click activates the shutter on the transmitted UBS
mode, ensuring the presence of the higher, i.e. correctly
injected, pulses. This scheme has the peculiar property of
selecting an invariant region of the Fock space with respect to
rotations of the polarization basis. As said, the action of the ID
device allows to decrease the noise due to the vacuum injection
into the amplifier since it preserves only the higher pulses, and hence
the ones that, with an higher probability, belong to the
amplification process. These considerations can be quantified in the
following way. The parameter of interest is the conditional injection probability,
i.e. the injection probability conditioned to the activation of the shutter given
by the threshold condition of the ID. We then evaluated numerically
this quantity for several values of the un-conditioned injection
probability $p$. It turns out that the value of $p_{cond}$ is
increased as shown in Fig.\ref{fig:TD_soglia}, in which we report
the trend of the conditional injection probability $p_{cond}$ as a function of the ID
threshold $h$.

\section{Deterministic transmitted state identification}
\label{sec:deterministic identification}

In this section we are interested in exploiting the action of a
different pre-selection strategy, no more based on the intensity
filtering but on a comparison between orthogonally polarized
signals. 
\begin{figure}[ht!]
\includegraphics[width=0.5\textwidth]{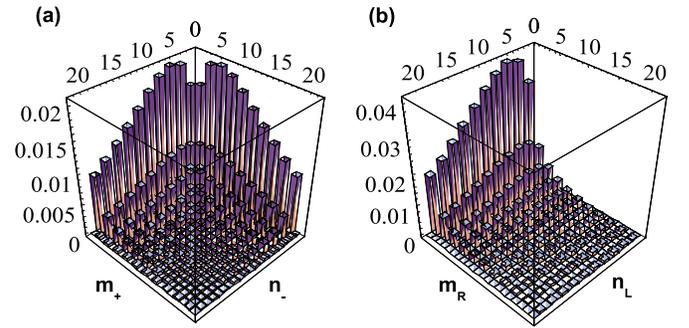}
 \caption{(Color online) (a) Probability distribution of the state $\ket{\Phi^{R}}$ as a function of the
 number of photons $\{\vec{\pi}_{+},\vec{\pi}_{-}\}$ . (b) Probability distribution of the state
 $\ket{\Phi^{R}}$ as a function of the  number of photons $\{\vec{\pi}_{R},\vec{\pi}_{L}\}$.
 In both distributions $g=1.5$.}
\label{fig:distributions}
\end{figure}
\begin{figure}[ht!]
\includegraphics[width=0.5\textwidth]{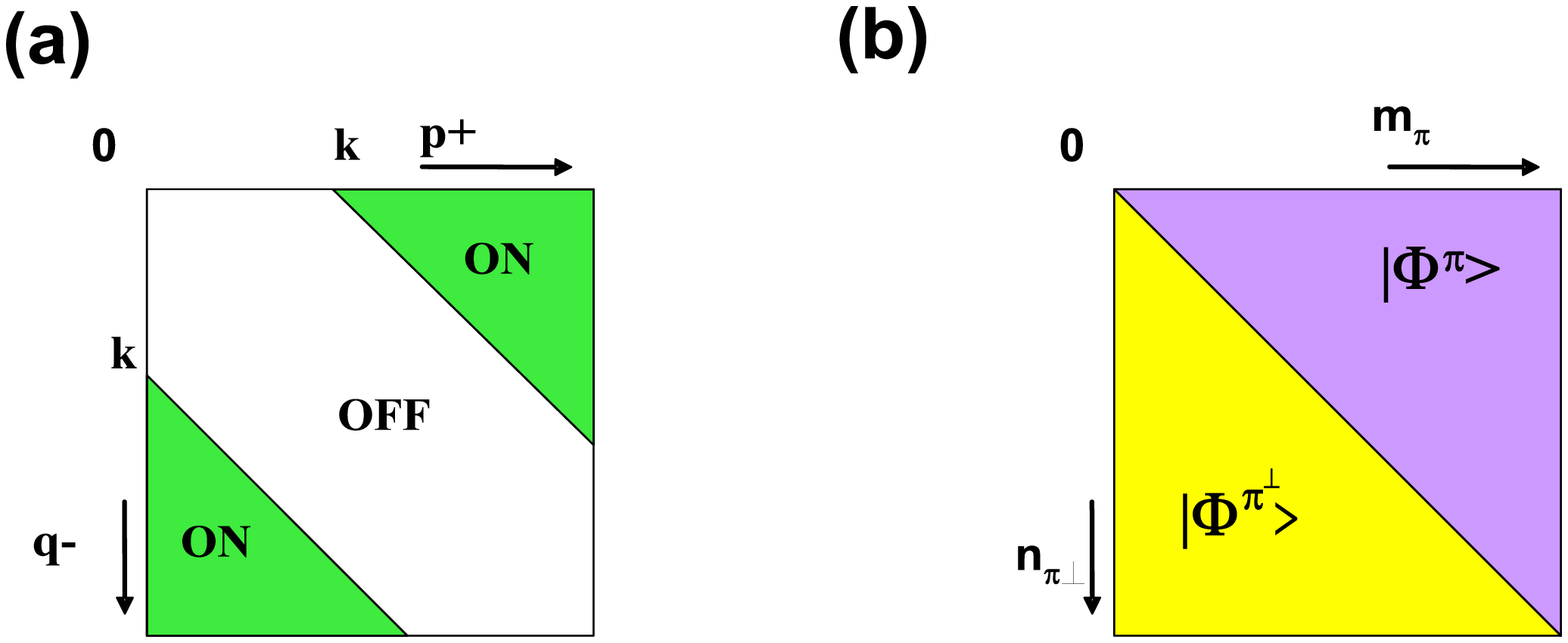}
 \caption{(Color online) (a) Measurement scheme adopted for the conditional activation of the shutter: if the OF, on the reflected mode,
  measures the state on the green regions, the shutter, on the transmitted mode, is conditionally activated. The green
  regions correspond to the state for which the signals belonging to orthogonal polarizations are unbalanced over a
  certain threshold $k$, i.e. $|p-q|\geq k$. (b) Scheme for the final detection of the output state: Conditioned on a measurement
  result in the ON region on the reflected mode, the transmitted mode is identified by a dichotomic measurement in the $\{\pi,\pi_{\perp}\}$ 
  basis. The diagonal contribution to the quantum state is assigned randomly to the state $\ket{\Phi^{\pi}}$ or $\ket{\Phi^{\pi\perp}}$.}
\label{fig:shutter activation}
\end{figure}
This configuration is illustrated in
fig.\ref{fig:conceptual_scheme}-II and is based on a peculiar
feature of the equatorial macro states. Indeed, any
macro-qubit belonging to the injection of an equatorial qubit,
due to the phase covariance of the amplifier, can be discriminated
through a measurement based on a comparison. Precisely, we can
measure the intensity signals belonging to orthogonal polarization
components of the same macro-state and subsequently compare them above a
certain threshold $k$. If analyzed in the same polarization basis
of the injected qubit, the two signals will be unbalanced with an high
probability. This can be explained by analyzing the
probability distribution of the amplified states, reported in
figure \ref{fig:distributions}: (a) in the mutually unbiased
equatorial polarization basis with respect to the injected state 
and (b) in the same basis as the injected qubit one.

We will address two cases in which the state generated by the
amplifier is either $\ket{\Phi^{+}}$ or $\ket{\Phi^{R}}$, obtained
by the amplification of a single photon polarized
$\vec{\pi}_{+}=\frac{\vec{\pi}_{H}+\vec{\pi}_{V}}{\sqrt{2}}$ and
$\vec{\pi}_{R}=\frac{\vec{\pi}_{H}+\imath \vec{\pi}_{V}}{\sqrt{2}}$,
respectively. In both cases the analysis basis corresponding to
the UBS reflected mode is fixed to
$\{\vec{\pi}_{+},\vec{\pi}_{-}\}$, while the transmitted mode is
analyzed in the same basis in which the injected qubit
has been encoded. Let us discuss the experimental setup shown in
Figure \ref{fig:conceptual_scheme}-II. The macro-state
$\ket{\Phi^{+}}$ (or $\ket{\Phi^{R}}$) generated by the QIOPA
impinges on the UBS. A small portion of the field is reflected on
mode $\mathbf{k_{d}}$ and measured on the
$\{\vec{\pi}_{+},\vec{\pi}_{-}\}$ basis. The two signals belonging
to orthogonal polarizations are then compared by an
orthogonality filter (OF). When the two signals are unbalanced,
i.e. $|p-q|>k$, being $p,q$ the number of photons
$\vec{\pi}_{+},\vec{\pi}_{-}$ polarized and $k$ an appropriate
threshold, the shutter on mode $\mathbf{k_{c}}$ is activated and
the field on that mode is conditionally transmitted (see
Fig.\ref{fig:shutter activation}). The macro-state
$\ket{\Phi^{+}}$ ($\ket{\Phi^{R}}$) is then analyzed in the
$\{\vec{\pi}_{+},\vec{\pi}_{-}\}$ (or
$\{\vec{\pi}_{R},\vec{\pi}_{L}\}$ ) basis. In the following
sections we will address the problem of discriminating the final
macro-state, given the acquired information on the small portion
of the reflected field.
%

\begin{figure}[ht!]
\includegraphics[width=0.5\textwidth]{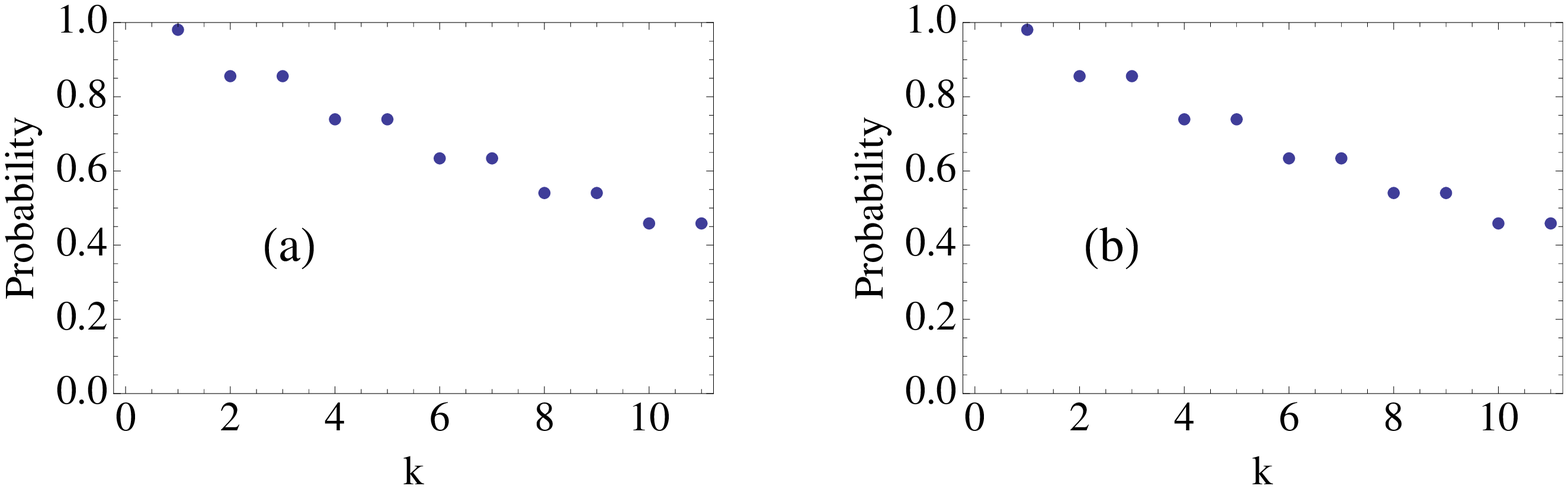}
 \caption{(Color online) (a) Probability of activating the shutter when the state $\ket{\Phi^{R}}$ is analyzed
 in the $\{\vec{\pi}_{+},\vec{\pi}_{-}\}$ basis versus the threshold k of the OF. (b) Probability of
 activating the shutter when the state $\ket{\Phi^{+}}$ is analyzed on the $\{\vec{\pi}_{+},\vec{\pi}_{-}\}$
 basis. The nonlinear gain of the amplifier is set at $g=1.5$.}
\label{fig:probability}
\end{figure}

%

\subsection{Probability of shutter activation}
Let us first evaluate the probability $P$ of activating the
shutter when the impinging state is detected on the
$\{\vec{\pi}_{+},\vec{\pi}_{-}\}$ basis, depending on the value of
$k$, with an OF technique.
As shown in figure \ref{fig:probability}, the probability of
activating the shutter is the same for the two output fields
$\ket{\Phi^{+}}$ and $\ket{\Phi^{R}}$. This result can be
explained by considering the probability distributions of the
state $\ket{\Phi^{R}}$ in the two mutually unbiased equatorial bases shown in figure
\ref{fig:distributions}. Due to the linearity of the quantum
mechanics, the state $\ket{\Phi^{R}}$ can be written as
$\ket{\Phi^{R}}=\frac{1}{\sqrt{2}}\left(
\ket{\Phi^{+}}+\imath \ket{\Phi^{-}}\right)$. Hence, due to the peculiar
features of the two macro-states $\ket{\Phi^{\pm}}$, that have
non-zero contributions for terms with different parity, the
probability distribution of the macro-state $\ket{\Phi^{R}}$  in
the $\{\vec{\pi}_{+},\vec{\pi}_{-}\}$ basis is given as the sum of
the two probability distributions of the states $\ket{\Phi^{+}}$
and $\ket{\Phi^{-}}$ in the same basis. 
Since shot by shot the OF
identifies the state $\ket{\Phi^{+}}$ or $\ket{\Phi^{-}}$ with the
same probability, the activation of the shutter has the same
probability of occurrence for any linear combination of
$\ket{\Phi^{-}}$ and $\ket{\Phi^{+}}$ impinging on the BS.

\subsection{Analysis of the Macro-state $\ket{\Phi^{+}}$}

Let us analyze the evolution of the state $\ket{\Phi^{+}}$ passing
through an unbalanced beam-splitter (UBS). We start with the
expression of the macro-qubit \cite{DeMa08}:

\begin{equation}
\begin{aligned}
\label{eq:Phi_piu}
\ket{\Phi^{+}}&=\frac{1}{C^{2}}\sum_{ij}^{\infty}\left(\frac{-\Gamma}{2}\right)^{j}\left(\frac{\Gamma}{2}\right)^{i}
\frac{\sqrt{(2j)!(2i+1)!}}{j!i!} \\
&\times \ket{(2i+1)+,2j-}_{b} =\\
&=\frac{1}{C^{2}}\sum_{ij}^{\infty}\left(\frac{-\Gamma}{2}\right)^{j}\left(\frac{\Gamma}{2}\right)^{i}
\frac{\mathbf{(b_{+}^{\dag})^{2i+1}}}{j!}\frac{\mathbf{(b_{-}^{\dag})^{2j}}}{i!}\ket{0}
\end{aligned}
\end{equation}

\noindent The UBS transformation equations for the creation
operators on spatial mode $\mathbf{b}$ read:

\begin{equation}
\mathbf{b_{\pm}}=\sqrt{\tau}\mathbf{c_{\pm}}+\imath \sqrt{1-\tau}\,\mathbf{d_{\pm}}
\end{equation}

\noindent where the subscript $\pm$ refers to the polarization
modes $\vec{\pi}_{\pm}=\frac{\vec{\pi}_{H}\pm \vec{\pi}_{V}}{\sqrt{2}}$, and
$\mathbf{c^{\dag}}$ and $\mathbf{d^{\dag}}$ refer to the creation
operators on the spatial modes transmitted and reflected by the
UBS. Hence after the UBS the output state becomes:

\begin{equation}
\begin{aligned}
\label{eq:Phi_piu_out_tmp}
\ket{\Phi^{+}}^{out}&=\frac{1}{C^{2}}\sum_{ij}^{\infty}\sum_{k}^{2i+1}\sum_{l}^{2j}\begin{pmatrix} 2i+1 \\
k\end{pmatrix}\begin{pmatrix} 2j \\
l\end{pmatrix}\left(\frac{-\Gamma}{2}\right)^{j}\\
&\; \left(\frac{\Gamma}{2}\right)^{i}
\frac{1}{j!}\frac{1}{i!}\left(\sqrt{\tau}\mathbf{c_{+}^{\dag}}\right)^{k}\left(\imath\sqrt{1-\tau}\mathbf{d_{+}^{\dag}}\right)^{2i+1-k}
\\ &\; \left(\sqrt{\tau}\mathbf{c_{-}^{\dag}}\right)^{l}
\left(\imath\sqrt{1-\tau}\mathbf{d_{-}^{\dag}}\right)^{2j-l}\ket{0} 
\end{aligned}
\end{equation}
By applying the creation operators to the vacuum state the output state reads:
\begin{equation}
\begin{aligned}
\label{eq:Phi_piu_out}
\ket{\Phi^{+}}^{out}&=\frac{1}{C^{2}}\sum_{ij}^{\infty}\sum_{k}^{2i+1}\sum_{l}^{2j}
\left(\frac{-\Gamma}{2}\right)^{j}\left(\frac{\Gamma}{2}\right)^{i}
\frac{1}{j!}\frac{1}{i!}\frac{\sqrt{\tau}^{k+l}}{\sqrt{k!}}\\
&\;\frac{(2i+1)!(2j)!}{\sqrt{(2i+1-k)!(2j-l)!}}\frac{(\imath \sqrt{1-\tau})^{2i+1+2j-k-l}}{\sqrt{l!}}\\
&\; \ket{k+,l-}_{d}\ket{(2i+1-k)+,(2j-l)-}_{c}
\end{aligned}
\end{equation}

Let us now consider the case in which the reflected mode by the
UBS is measured on the $\{\vec{\pi}_{+},\vec{\pi}_{-}\}$ basis.
The state $\ket{p+,q-}_{d}$ is detected on the reflected mode, the
transmitted state then reads:

\begin{eqnarray}
\ket{\Phi^{+}}^{meas}&=&\frac{1}{C^{2}}\sum_{ij}^{\infty}\left(\frac{-\Gamma}{2}\right)^{j}\left(\frac{\Gamma}{2}\right)^{i}
\frac{\sqrt{\tau}^{2i+1+2j-p-q}}{j!} \nonumber \\
&\;& \frac{(\imath \sqrt{1-\tau})^{p+q}}{i!}
\frac{(2i+1)!(2j)!}{\sqrt{p!q!(2i+1-p)!(2j-q)}} \nonumber \\&\;&
\ket{(2i+1-p)+,(2j-q)-}_{c}
\end{eqnarray}

We are interested in investigating the distinguishability between 
orthogonal macro-states by varying the pre-selection performed over the 
multiphoton state itself. It is worth noting that our  investigation addresses 
the coherence between two different polarization modes which is quantified 
by the visibility of the intensity curve obtained by rotating a polarizer.
Our first scope is to investigate the visibility of the transmitted mode
as a function of the unbalancement between $\vec{\pi}_{+}$ and
$\vec{\pi}_{-}$ photons, detected on the reflected mode. 
Namely, if $|p-q|>k$ on mode $\mathbf{k_{d}}$, what is the visibility of
the state $\ket{\Phi^{+}}^{meas}$ on mode $\mathbf{k_{c}}$?\\
This quantity can be quantified in the following way. Due to the
peculiar shape of the photon number probability distribution
(figure \ref{fig:distributions}-(b)), the identification of the
state $\ket{\Phi^{+}}$ can be performed by discriminating between
the number of photons $\vec{\pi}_{+}$ and $\vec{\pi}_{-}$
polarized. Let us define the following quantities:
$P^{+}(m,n|p,q)$ is the probability that, if the state
$\ket{p+,q-}_{d}$ is detected on spatial mode $\mathbf{k_{d}}$,
$m>n$ is obtained on spatial mode $\mathbf{k_{c}}$, and hence the
macro-state $\ket{\Phi^{+}}$ is identified ($m,n$ being the number
of photons $\vec{\pi}_{+}$ and $\vec{\pi}_{-}$ polarized). On the
contrary $P^{-}(m,n|p,q)$ is the probability that, given the
detection of the state $\ket{p+,q-}_{d}$ on spatial mode
$\mathbf{k_{d}}$, $n>m$ is obtained on spatial mode
$\mathbf{k_{c}}$, and hence the macro-state $\ket{\Phi^{-}}$ is
identified, even if the initial state impinging on the UBS was $\ket{\Phi^{+}}$.
We can than derive the visibility as a function of the threshold
$k$ such that $\vert p-q \vert>k$:

\begin{equation}
V(k)=\frac{\sum_{m,n}\sum_{p,q}\left(P^{p,q \; +}_{m,n}(k)-
P^{p,q \; -}_{m,n}(k) \right)}{\sum_{m,n}\sum_{p,q}
\left(P^{p,q \; +}_{m,n}(k)+P^{p,q \; -}_{m,n}(k)\right)}
\end{equation}

\noindent where $P^{p,q \; \pm}_{m,n}(k) = P^{\pm}(m,n|\vert p-q \vert >k)$. 
The trend of visibility as a function of $k$ is
reported on Figure \ref{fig:visibility}-(a). We observe that,
increasing the value of $k$, hence detecting a higher
unbalancement between $\vec{\pi}_{+}$ and $\vec{\pi}_{-}$ photons
on mode $\mathbf{k_{d}}$, we obtain an higher visibility of the
state on mode $\mathbf{k_{c}}$.

\subsection{Analysis of the Macro-state $\ket{\Phi^{R}}$}

Let us consider the case in which the state $\ket{\Phi^{R}}$
impinges on the UBS:

\begin{equation}
\begin{aligned}
\label{eq:Phi_R}
\ket{\Phi^{R}}&=\frac{1}{C^{2}}\sum_{ij}^{\infty}\left(\frac{\imath \Gamma}{2}\right)^{j}\left(\frac{\imath \Gamma}{2}\right)^{i}
\frac{\sqrt{(2j)!}}{j!}\frac{\sqrt{(2i+1)!}}{i!} \\
&\times \ket{(2i+1)R,(2j)L}_{b} = \\
&=\frac{1}{C^{2}}\sum_{ij}^{\infty}\left(\frac{\imath\Gamma}{2}\right)^{j}\left(\frac{\imath\Gamma}{2}\right)^{i}
\frac{\mathbf{(b_{R}^{\dag})^{2i+1}}}{j!}\frac{\mathbf{(b_{L}^{\dag})^{2j}}}{i!}\ket{0}
\end{aligned}
\end{equation}

\noindent After the UBS the state can be written as:

\begin{equation}
\begin{aligned}
\label{eq:Phi_R_out}
\ket{\Phi^{R}}^{out}&=\frac{1}{C^{2}}\sum_{ij}^{\infty}\sum_{k}^{2i+1}\sum_{l}^{2j}\left(\frac{\imath\Gamma}{2}\right)^{j}\left(\frac{\imath\Gamma}{2}\right)^{i}
\frac{1}{j!}\frac{1}{i!} \frac{\sqrt{\tau}^{k+l}}{\sqrt{k!}}\\
&\;\frac{(2i+1)!(2j)!}{\sqrt{(2i+1-k)!(2j-l)!}}\frac{(\imath\sqrt{1-\tau})^{2i+1+2j-k-l}}{\sqrt{l!}}\\
&\;\ket{k R,l L}_{c}\ket{(2i+1-k)R,(2j-l)L}_{d}
\end{aligned}
\end{equation}

The state on mode $\mathbf{k_{d}}$ is then measured in the
$\{\vec{\pi}_{+},\vec{\pi}_{-}\}$ basis. The state
$\ket{(2i+1-k)R,(2j-l)L}_{d}$ can then be rewritten as:

\begin{equation}
\begin{aligned}
&\;\ket{(2i+1-k)R,(2j-l)L}=\sum_{r}^{2i+1-k}\sum_{s}^{2j-l}\frac{1}{\sqrt{2}^{2i+1+2j-k-l}}\\
&\,\frac{1}{\sqrt{(2i+1-k)!}}\frac{1}{\sqrt{(2j-l)}}\begin{pmatrix}2i+1-k\\s\end{pmatrix}\begin{pmatrix}2j-l\\s\end{pmatrix}\times\\
&\,\sqrt{(s+r)!(2i+1+2j-k-l-s-r)!}\imath^{2i+1-k-r}\imath^{2j-l-s}\\
&\,\ket{(r+s)+,(2i+1+2j-k-l-r-s)-}_{d}
\end{aligned}
\end{equation}

\noindent and the overall state reads:

\begin{widetext}

\begin{eqnarray}
&\,&\ket{\Phi^{R}}^{out}=\frac{1}{C^{2}}\sum_{ij}^{\infty}\sum_{k}^{2i+1}\sum_{l}^{2j}\sum_{r}^{2i+1-k}\sum_{s}^{2j-l}
\left(\frac{\imath\Gamma}{2}\right)^{j}\left(\frac{\imath\Gamma}{2}\right)^{i}\frac{1}{i!}\frac{1}{j!}
\frac{(2i+1)!(2j)!\imath^{2i+1-k-r}(-\imath)^{2j-l-s}}{\sqrt{(2i+1-k)!(2j-l)!k!l!}}\frac{\sqrt{\tau}^{k+l}\sqrt{\imath(1-\tau)}^{2i+1+2j-k-l}}{\sqrt{2}^{2i+1+2j-k-l}}
\nonumber\\
&\;&\frac{\sqrt{(r+s)!}
\sqrt{(2i+1+2j-k-l-r-s)!}}{\sqrt{(2i+1-k)!(2j-l)!}}\begin{pmatrix}2i+1-k\\r\end{pmatrix}\begin{pmatrix}2j-l\\s\end{pmatrix} 
\; \ket{(r+s)+,(2i+1+2j-k-l-r-s)-}_{d}\ket{k R,l L}_{c}
\end{eqnarray}

\end{widetext}

\noindent If the state on mode $\mathbf{k_{d}}$ is detected :
$\ket{(r+s)+,(2i+1+2j-k-l-r-s)-}_{d}=\ket{p+,q-}_{d}$, the state
on mode $\mathbf{k_{c}}$ is:

\begin{widetext}

\begin{eqnarray}\label{eq:Phi_R_meas}
&\,&\ket{\Phi^{R}}^{meas}=\frac{1}{C^{2}}\sum_{ij}^{\infty}\sum_{l}^{2j}\sum_{s}^{2j-l}
\left(\frac{\imath\Gamma}{2}\right)^{j}\left(\frac{\imath\Gamma}{2}\right)^{i}\frac{1}{i!}\frac{1}{j!}
\frac{(2i+1)!(2j)!}{\sqrt{(l+p+q-2j)!(2j-l)!}}\frac{\sqrt{\tau}^{2i+2j+1-p-q}}{\sqrt{(2i+1+2j-l-p-q)!l!}}\nonumber\\
&\;&
\frac{\sqrt{p!q!}}{\sqrt{2}^{p+q}}\frac{\sqrt{\imath(1-\tau)}^{p+q}}{\sqrt{(l+p+q-2j)!(2j-l)!}}\begin{pmatrix}l+p+q-2j\\p-s\end{pmatrix}
\begin{pmatrix}2j-l\\s\end{pmatrix}\imath^{l+q-2j}(-\imath)^{2j-l-s}\ket{(2i+1+2j-l-p-q)R,lL}_{c}\nonumber\\
\end{eqnarray}

\end{widetext}

\noindent where the following conditions have to be satisfied:
\begin{eqnarray}
p&>&s \nonumber\\
2i+1+2j&>& l+p+q \\
2j&<& l+p+q \nonumber
\end{eqnarray}

\noindent If the state (\ref{eq:Phi_R_meas}) is measured in the
polarization basis $\{\vec{\pi}_{R},\vec{\pi}_{L}\}$ obtaining a
state $\ket{mR,nL}_{c}$, the corresponding probability amplitude
is:

\begin{equation}
\begin{aligned}
\label{eq:amplitude_R}
&\,\frac{1}{C^{2}}\sum_{j}^{\infty}\sum_{s}^{2j-n}
\left(\frac{\imath\Gamma}{2}\right)^{j}\left(\frac{\imath\Gamma}{2}\right)^{(-2j+m+n+p+q-1)/2} \frac{(-\imath)^{2j-n-s}}{\sqrt{m!n!}} \\
&\, \frac{1}{j!} \frac{1}{\left(\frac{-2j+m+n+p+q-1}{2}\right)!} \frac{(-2j+p+q+m+n)!(2j)!}{\sqrt{(n+q-2j+s)!(p-s)!}}\\
&\, \frac{\sqrt{p!q!}}{(2j-n-s)!s!} \sqrt{\tau}^{m+n}(\imath\sqrt{1-\tau})^{p+q}\frac{1}{\sqrt{2}^{p+q}}\imath^{n+p+q-2j-r}
\end{aligned}
\end{equation}

\noindent and the probability of measuring the state
$\ket{mR,nL}_{c}$ is given by:

\begin{equation}
\begin{aligned}
&\,P(m,n|p,q)=\frac{1}{C^{4}}\sum_{j=0}^{\infty}\sum_{s}^{2j-n}\sum_{i}^{\infty}\sum_{r}^{2i-n}\frac{1}{C^{4}}\left(\frac{\Gamma}{2}\right)^{m+n+p+q-1}\\
&\, \frac{p!q!}{m!n!}\frac{(-1)^{r-s}}{2^{p+q}}
\frac{\tau^{m+n}(1-\tau)^{p+q}}{\left(\frac{p+q+m+n-2j-1}{2}\right)!j!}
\frac{1}{\left(\frac{p+q+m+n-2i-1}{2}\right)!i!}\\
&\,
\frac{(p+q+m+n-2j)!(2j)!}{(n+q+s-2j)!s!(2j-n-s)!(p-s)!}\times\\
&\,
\frac{(p+q+m+n-2i)!(2i)!}{(q+n+r-2i)!r!(2i-n-r)!(p-r)!}
\end{aligned}
\end{equation}

\noindent The visibility of the macro-state reads:

\begin{equation}
V(k)=\frac{\sum_{m,n}\sum_{p,q}\left(P^{p,q \; R}_{m,n}(k)-
P^{p,q \; L}_{m,n}(k) \right)}{\sum_{m,n}\sum_{p,q}
\left(P^{p,q \; R}_{m,n}(k)+P^{p,q \; L}_{m,n}(k)\right)}
\end{equation}

\noindent where $P^{p,q \; R,L}_{m,n}(k) = P^{R,L}(m,n|\vert p-q \vert >k)$.
Here, $P^{R}(m,n|\vert p-q \vert >k)$ is the probability
that, given the detection of the state $\ket{p+,q-}_{d}$ on mode
$\mathbf{k_{c}}$, $m>n$ is obtained on mode $\mathbf{k_{d}}$
($m$($n$), number of photons polarized
$\vec{\pi}_{R}$($\vec{\pi}_{L}$)). In this case the state
$\ket{\Phi^{R}}$ is identified; conversely the state
$\ket{\Phi^{L}}$ is detected even if the state
$\ket{\Phi^{R}}$ impinged on the UBS.

\begin{figure}[ht!]
\centering
\includegraphics[width=0.45\textwidth]{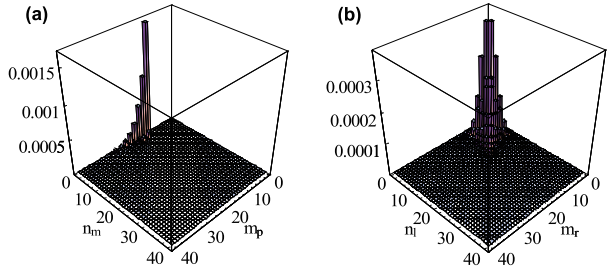}
\caption{(Color online) Selected region for the $\vert \Phi^{+} \rangle$ state after the measurement with an 
OF in the $\{\vec{\pi}_{+}, \vec{\pi}_{-} \}$ basis. (a) Photon number distribution in the 
$\{\vec{\pi}_{+}, \vec{\pi}_{-} \}$ basis. (b) Photon number distribution in the 
$\{\vec{\pi}_{R}, \vec{\pi}_{L} \}$ basis. In both cases $k=10$ and $g=1.2$.}
\label{fig:OF_filtering_basis}
\end{figure}

We observe that the visibility of the state $\ket{\Phi^{R}}$ in
the case in which a small portion of the overall state is measured
on the $\{\vec{\pi}_{+},\vec{\pi}_{-}\}$ polarization basis, is a
decreasing function of the threshold $k$ ($|p-q|>k$). This trend
is shown in figure \ref{fig:visibility}-(c). The decreasing trend
of visibility can be explained by considering that the
measurements in the two polarization basis correspond to two
non-commuting operators acting on the same initial state. Indeed,
for asymptotically high values of the threshold $k \rightarrow
\infty$, the measurement of the $\hat{\Pi}_{i}$ operators that
describe the OF tends to the measurement of the pseudo-spin
operators $\hat{\Sigma}_{i}$: i.e
$\hat{\Sigma}_{1}=\ket{\Phi^{+}}\bra{\Phi^{+}}-\ket{\Phi^{-}}\bra{\Phi^{-}}$
or
$\hat{\Sigma}_{2}=\ket{\Phi^{R}}\bra{\Phi^{R}}-\ket{\Phi^{L}}\bra{\Phi^{L}}$.
More details on the relationship between the OF device and the
pseudo-Pauli operator can be found in Ref.\cite{Spag10}. In view
of this consideration, the measurement on the $\mathbf{k}_{C}$
mode corresponds to the measurement of the $\hat{\Sigma}_{i}$
operators. The information gained on this mode about one of the
two pseudo-spin operator acting on the macro qubit does not allow
to gain information about orthogonal pseudo-spin operator. As a
further remark, let us stress that this feature of the OF
measurement is related to the filtering of different regions of
the Fock space depending on the analyzed basis. The portion of the
state that survives the action of the OF is indeed different if
measured on the $\{\vec{\pi}_{+},\vec{\pi}_{-}\}$ basis or in the
$\{\vec{\pi}_{R},\vec{\pi}_{L}\}$ one and is shown in
figure \ref{fig:OF_filtering_basis}. 

\section{Probabilistic transmitted state identification}
\label{sec:probabilistic identification}

In the previous sections we have shown how the visibility of the
macro qubit obtained by a pure dichotomic measurement can be
modified if a small portion of the beam is identified by a
probabilistic measurement strategy. 
%

\begin{figure}[ht!]
\centering
\includegraphics[width=0.5\textwidth]{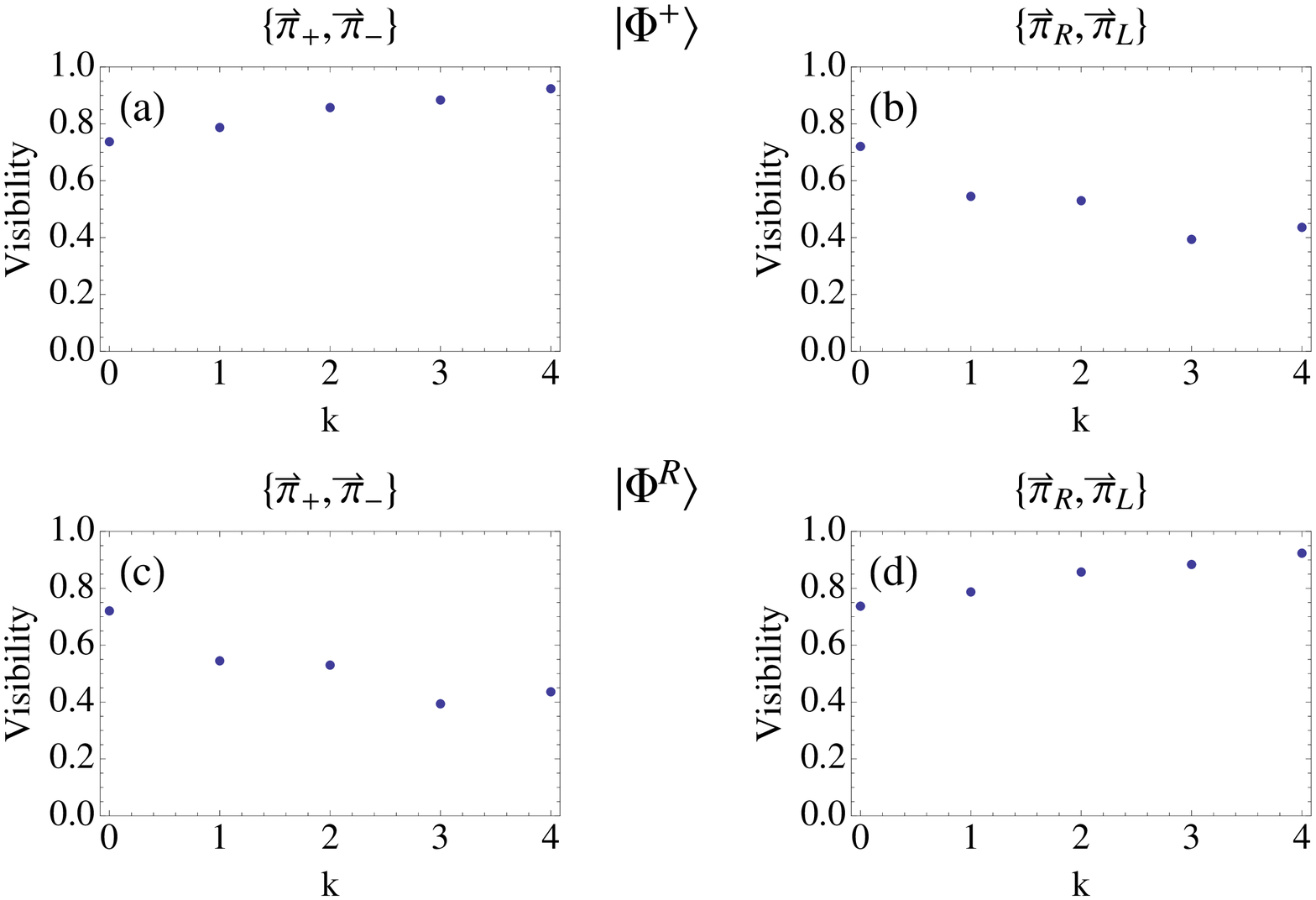}
\caption{(Color online) (a)-(b)Trend of the visibility of the state
$\ket{\Phi^{+}}$ measured in the basis
$\{\vec{\pi}_{+},\vec{\pi}_{-}\}$ and $\{\vec{\pi}_{R},\vec{\pi}_{L}\}$ respectively  as a function of the threshold
$k$. (c)-(d) Trend of the visibility of the state $\ket{\Phi^{R}}$
measured  in the basis $\{\vec{\pi}_{+},\vec{\pi}_{-}\}$ and $\{\vec{\pi}_{R},\vec{\pi}_{L}\}$ respectively as a
function of the threshold $k$. The numerical results have been
obtained for the value of the gain parameter $g=1.1$.}
\label{fig:visibility}
\end{figure}
%

\begin{figure}[ht!]
\centering
\includegraphics[width=0.45\textwidth]{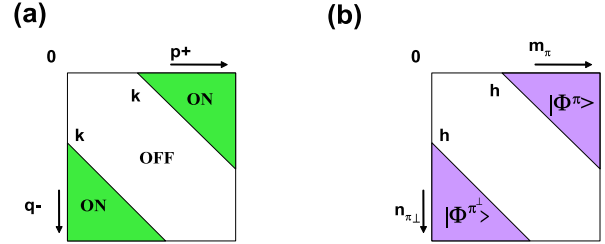}
 \caption{(Color online)(a) Conditional activation of the shutter: if the OF acting on the reflected mode
  measures the state on the green regions, the shutter, on the transmitted mode, is conditionally activated. The green
  regions correspond to the state for which the signals belonging to orthogonal polarizations are unbalanced over a
  certain threshold $k$, i.e. $|p-q|\geq k$. (b) Corresponding to the ON region on the reflected mode, the transmitted mode is
  identified by a probabilistic measurement in the $\{\pi,\pi_{\perp}\}$ basis. The identification condition is $|m-n|\geq h$. }
\label{fig:shutter activation_double_threshold}
\end{figure}

This section addresses the trend of the macro-states
visibility when the field is split in two equal parts by a $0.5/0.5$
beam-splitter and both the reflected and the transmitted states are
detected through the OF device. In this case the measurement
schemes are shown in figures \ref{fig:conceptual_scheme}-(III) and
\ref{fig:shutter activation_double_threshold}: the
OF technique is applied in order to extract the maximum
information available from the two measured states.

\begin{figure}[ht!]
\centering
\includegraphics[width=0.45\textwidth]{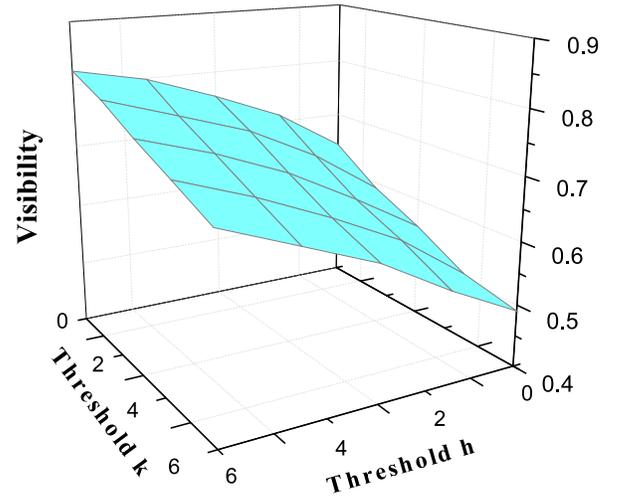}
 \caption{(Color online) Trend of the visibility of the state $\ket{\Phi^{R}}$ for different values of the threshold $h$ on the transmitted mode and of the threshold $k$ on the reflected one. The numerical
 result has been obtained for a value of the parameter $g=1.2$.}
\label{fig:OF_50_50}
\end{figure}

\begin{figure}[ht!]
\centering
\includegraphics[width=0.45\textwidth]{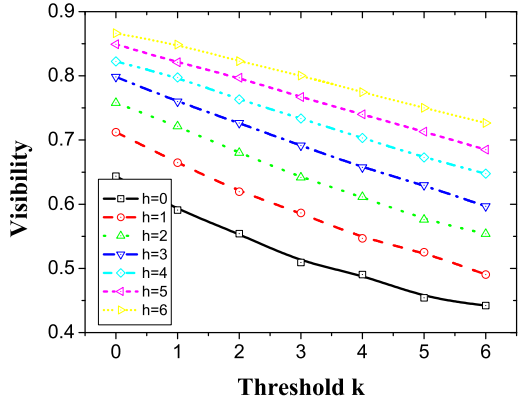}
 \caption{(Color online) Trend of the macro-state visibility as a function of the threshold $k$ on the reflected mode, fixed the threshold $h$ on the transmitted one.}
\label{fig:visibility_50_50}
\end{figure}

We consider the case in which the portion on the reflected mode is
analyzed in the polarization basis orthogonal to the codification
one. In figure \ref{fig:OF_50_50} is reported the trend of
visibility as a function of the threshold $h$ on the transmitted
mode, and $k$ on the reflected one. The two polarization analysis
basis are chosen to be mutually unbiased. It can be seen that for
equal values of the two thresholds  $h=k$ the visibility reaches a
value around $0.64$, the same obtained through a pure dichotomic
measurement, without any pre-selection procedure on the
macro-state. In figure \ref{fig:visibility_50_50} is reported the
trend of the visibility as a function of the threshold on the
reflected mode, keeping fixed the value of the threshold on the
transmitted one.  We can see that the visibility of the
transmitted state decreases when the threshold on the reflected
mode increases. If the threshold on the transmitted mode is
greater than the one on the reflected mode, the visibility results
to be higher than $0.64$, as expected by the action of the OF,
which allows a better discrimination of the macro-state, measured
in the codification polarization basis. Otherwise it can be seen
how, decreasing the threshold $h$ below the threshold $k$, the
visibility decreases below the ``no filtering value''.

From the analysis performed in this paper we can conclude that the macro states are not suitable for quantum cryptography. The action on a portion of the state can indeed be seen as an eavesdropping attack. If the state is measured in the codification basis, the visibility of the final state results to increase as shown in figures \ref{fig:visibility} (a)-(d). This means that the conclusive results for the eavesdropper would coincide with the conclusive results for the receiver, and the eavesdropper can gain information on the macrostates without introducing noise. Otherwise if the state is measured by the eavesdropper in the wrong basis, the visibility at the receiver is not affected if the state is measured above a certain filtering threshold.
According to these considerations, an eavesdropper could then develop a strategy in which he measures its part of the transmitted state in two bases. With this approach he could gain information on the transmitted signal by considering only the measurement outcome in the right basis, and only a small amount of noise is introduced by keeping the filtering thresholds above a certain value.
Related to the security of the macro-states is the possibility of performing a non-locality tests upon them. As a final remark for this section, we remind that the adoption of the OF device at the measurement stage is not suitable for a non-locality test, since the filtered portion of the state is \textit{dependent} on the measurement basis \cite{Vite10}.
We will then address the non-locality task in the following section.

\section{Pre-selection for entanglement and non-locality
tests}\label{sec:non-locality tests}

In this section we shall investigate a pre-selection scheme
based on a conditional operation driven by the measurement of a
portion of the multiphoton state in two different polarization basis.
The setup of this pre-selection scheme is reported in fig.\ref{fig:conceptual_scheme}-IV.
A small portion of the generated multiphoton state is reflected by an unbalanced beam-splitter
of transmittivity $T=0.9$ and subsequently split by a 50/50 beam-splitter in two equal
parts. One of the two parts is measured in an equatorial $\{ \vec{\pi}_{\beta},
\vec{\pi}_{\beta_{\bot}} \}$ basis by two photomultipliers, and the photocurrents
$\{ I_{\beta}, I_{\beta_{\bot}}\}$ are analyzed by an OF device
[Fig.\ref{fig:shutter activation}]. The other part undergoes the same measurement
process in a different equatorial basis $\{ \vec{\pi}_{\beta'}, \vec{\pi}_{\beta'_{\bot}} \}$.

When the threshold condition $\vert I_{\pi} - I_{\pi_{\bot}} \vert > k$ [Fig.\ref{fig:shutter activation}]
is realized in both branches, measured respectively in the polarization basis
$\{ \vec{\pi}_{\beta}, \vec{\pi}_{\beta_{\bot}} \}$ and
$\{ \vec{\pi}_{\beta^{\prime}}, \vec{\pi}_{\beta^{\prime}_{\bot}} \}$,
a TTL electronic signal is sent to conditionally activate the optical shutter placed in
the optical path of the remaining part of the multiphoton state. Then, the field is analyzed at the
measurement stage with the dichotomic strategy discussed in the previous paragraphs.
For this pre-selection method, the relevant parameter is the angle $\phi$ between the two bases
$\{ \vec{\pi}_{\beta}, \vec{\pi}_{\beta_{\bot}} \}$ and $\{ \vec{\pi}_{\beta^{\prime}},
\vec{\pi}_{\beta^{\prime}_{\bot}} \}$ in which the small portion of the beam is analyzed.
The angle $\phi$ is defined according to the relations between the two polarization bases:
\begin{eqnarray}
\vec{\pi}_{\beta'} &=& e^{\imath \frac{\phi}{2}} \left[ \cos \left( \frac{\phi}{2} \right) \vec{\pi}_{\beta} - \imath
\sin \left( \frac{\phi}{2} \right) \vec{\pi}_{\beta_{\bot}} \right] \\
\vec{\pi}_{\beta'_{\bot}} &=& e^{\imath \frac{\phi}{2}} \left[ - \imath \sin \left( \frac{\phi}{2} \right)
\vec{\pi}_{\beta} + \cos \left( \frac{\phi}{2} \right) \vec{\pi}_{\beta_{\bot}} \right]
\end{eqnarray}

\begin{figure}[ht!]
\centering
\includegraphics[width=0.4\textwidth]{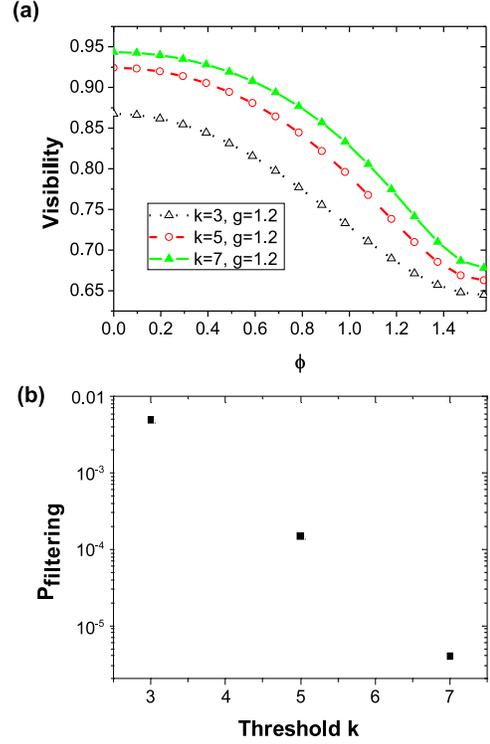}
\caption{(Color online) (a) Trend of the visibility for the double-filtering technique as a function of the angle $\phi$
between the two polarization bases $\{ \vec{\pi}_{\beta}, \vec{\pi}_{\beta_{\bot}} \}$ and $\{ \vec{\pi}_{\beta^{\prime}},
\vec{\pi}_{\beta^{\prime}_{\bot}} \}$ of the pre-selection measurement. Square black points correspond
to $k=3$, circular red points to $k=5$ and triangular green points to $k=7$. (b) Filtering probability
of the scheme as a function of the threshold $k$ at the pre-selection measurement stage. All graphs correspond to $g=1.2$.}
\label{fig:visibility_angle}
\end{figure}

Let us begin by analyzing the trend of the visibility of the fringe pattern obtained by varying
the equatorial polarization $\vec{\pi}_{\alpha}$ of the injected single-photon state
in the amplifier. More specifically, we analyze how the visibility changes as a function of the
angle $\phi$ between the two bases of the pre-selection branch. In Fig.\ref{fig:visibility_angle}
we show the numerical results obtained by calculating the visibility according to the standard definition
$V= \frac{I_{max}-I_{min}}{I_{max}+I_{min}}$. In this case, the visibility is evaluated according to
the following expression:

\begin{widetext}

\begin{equation}
\label{eq:visi_preselection}
V(k)=\frac{\sum_{m>n}P_{    \overline{\alpha}} \left[ m,n \big|(\vert I_{\beta} - I_{\beta_{\bot}} \vert > k) \cap (\vert I_{\beta'} - I_{\beta'_{\bot}} \vert > k) \right] - \sum_{m<n} P_{\overline{\alpha}} \left[ m,n \big|(\vert I_{\beta} - I_{\beta_{\bot}} \vert > k) \cap (\vert I_{\beta'} - I_{\beta'_{\bot}} \vert > k) \right] }{\sum_{m>n} P_{\overline{\alpha}} \left[ m,n \big|(\vert I_{\beta} - I_{\beta_{\bot}} \vert > k) \cap (\vert I_{\beta'} - I_{\beta'_{\bot}} \vert > k) \right] + \sum_{m<n} P_{\overline{\alpha}} \left[ m,n \big|(\vert I_{\beta} - I_{\beta_{\bot}} \vert > k) \cap (\vert I_{\beta'} - I_{\beta'_{\bot}} \vert > k) \right] }
\end{equation}

\noindent Here $P_{\overline{\alpha}} \left[ m,n \big|(\vert I_{\beta} - I_{\beta_{\bot}} \vert > k) \cap (\vert I_{\beta'} - I_{\beta'_{\bot}}
\vert > k) \right]$ is the photon-number distribution of the state $\vert \Phi^{\overline{\alpha}} \rangle$ after the
pre-selection stage. More specifically, the value of $\overline{\alpha}$ is chosen in order to maximize
the contribution of the $\sum_{m>n}$ term and minimize the contribution of the $\sum_{m<n}$ term:
\begin{eqnarray}
I_{max} &=& \sum_{m>n}P_{\overline{\alpha}} \left[ m,n \big|(\vert I_{\beta} - I_{\beta_{\bot}} \vert > k) \cap (\vert I_{\beta'} - I_{\beta'_{\bot}} \vert > k) \right]  \\
I_{min} &=& \sum_{m<n}P_{\overline{\alpha}} \left[ m,n \big|(\vert I_{\beta} - I_{\beta_{\bot}} \vert > k) \cap (\vert I_{\beta'} - I_{\beta'_{\bot}} \vert > k) \right]
\end{eqnarray}

\end{widetext}

Eq.(\ref{eq:visi_preselection}) then coincides with the usual definition of visibility.
We note that the visibility is higher for smaller angles $\phi$, since in that case a strong
projection of the state is performed in two close bases. This condition is equivalent to the scheme
of Fig.\ref{fig:conceptual_scheme}-(II), where the OF measurement performed in one basis
allows to obtain a better discrimination of the detected state only in the polarization
basis of the pre-selection measurement [Fig.\ref{fig:visibility} (a)-(b)]. When $\phi$ is high, a lower
visibility can be achieved since the projection of the macrostate occurs in two distant bases.
In this case, the increasing effect of the pre-selection in one basis on the visibility is in
contrast with the decreasing effect of the pre-selection in the other basis, as shown in Sec.III.

We conclude this section by discussing the feasibility of a non-locality test by exploiting the proposed
pre-selection method. We consider the case of a CHSH inequality \cite{Clau69}.
Let us briefly summarize in the light of a local hidden variable (LHV)
theory the content of Bell's inequalities for a set of dichotomic
observables. Consider a quantum state described by
the density matrix $\hat{\rho}$ defined in the Hilbert space $\mathcal{H}%
_{1}\otimes \mathcal{H}_{2}$. Define $\hat{O}_{a}^{i}$ the positive operator
acting on subspace $\mathcal{H}_{1}$, and the probability of finding the
value $i$ after the measurement $a$ as given by $\mathrm{Tr} \left[\hat{\rho}
(\hat{O}_{a}^{i}\otimes \hat{I}) \right]$.
The same relation holds for the positive operator $\hat{O}_{b}^{j}$ acting on
subspace $\mathcal{H}_{2}$.

The existence of a LHV model implies that the expectation values of the $a$
and $b$ observables are predetermined by the value of the parameter $\lambda $: \{$
X_{a}(\lambda),X_{a^{\prime }}(\lambda),X_{b}(\lambda),X_{b^{\prime }}(\lambda)\},$
hence the product $a\cdot b$ is equal to $X_{a}(\lambda )X_{b}(\lambda )$. For a
fixed value of $\lambda $ the variables $X_{n}$ with $n=\{a,b,a^{\prime },b^{\prime }\}$
take the values ${-1,1}$ and satisfy the CHSH inequality:

{\small
\begin{equation}
X_{a}(\lambda )X_{b}(\lambda )+X_{a}(\lambda )X_{b^{\prime }}(\lambda
)+X_{a^{\prime }}(\lambda )X_{b}(\lambda )-X_{a^{\prime }}(\lambda
)X_{b^{\prime }}(\lambda )\leq 2  \label{eq:random_inequality}
\end{equation}
} The same inequality holds by integrating this equation on the space of the
hidden variable $(\lambda )$:

\begin{eqnarray}
&\,&\int_{\Omega }d\mathbb{P}(\lambda )X_{a}(\lambda )X_{b}(\lambda
)+\int_{\Omega }d\mathbb{P}(\lambda )X_{a}(\lambda )X_{b^{\prime }}(\lambda
)+  \notag  \label{eq:random_inequality_integrated} \\
&\,&\int_{\Omega }d\mathbb{P}(\lambda )X_{a^{\prime }}(\lambda
)X_{b}(\lambda )-\int_{\Omega }d\mathbb{P}(\lambda )X_{a^{\prime }}(\lambda
)X_{b^{\prime }}(\lambda )\leq 2  \notag \\
&&
\end{eqnarray}%
where $\mathbb{P}(\lambda )$ is the measure of the $\lambda $ probability space. If
there is a local hidden variables model for quantum measurement taking
values $[-1,+1]$, then the following inequality must be satisfied:
\begin{equation}
S_{CHSH} = E^{\rho }(a,b)+E^{\rho }(a,b^{^{\prime }})+E^{\rho }(a^{^{\prime
}},b)-E^{\rho }(a^{^{\prime }},b^{^{\prime }})\leq 2
\label{eq:CHSH_inequality}
\end{equation}%
where $E^{\rho }(a,b)=\int_{\Omega }X_{a}(\lambda )X_{b}(\lambda )d\mathbb{P}%
(\lambda )$. The violation of (\ref{eq:CHSH_inequality}) proves that a LHV
variables model for the considered experiment is impossible.

We consider the case in which the angle $\phi$ between the two
bases $\{ \vec{\pi}_{\beta}, \vec{\pi}_{\beta_{\bot}} \}$ and $\{
\vec{\pi}_{\beta^{\prime}}, \vec{\pi}_{\beta^{\prime}_{\bot}} \}$
is set at $\phi = \pi/4$. This choice is motivated by the
following considerations. On one side, low values of $\phi$ would
lead to a micro-macro state possessing strong correlations only in
one polarization basis, thus not allowing to violate a Bell's
inequality. On the other side, high values of $\phi$ does not
allow to obtain the necessary enhancement in the correlations of
the micro-macro system to violate a Bell's inequality. 
\begin{figure*}[ht!]
\centering
\includegraphics[width=0.8\textwidth]{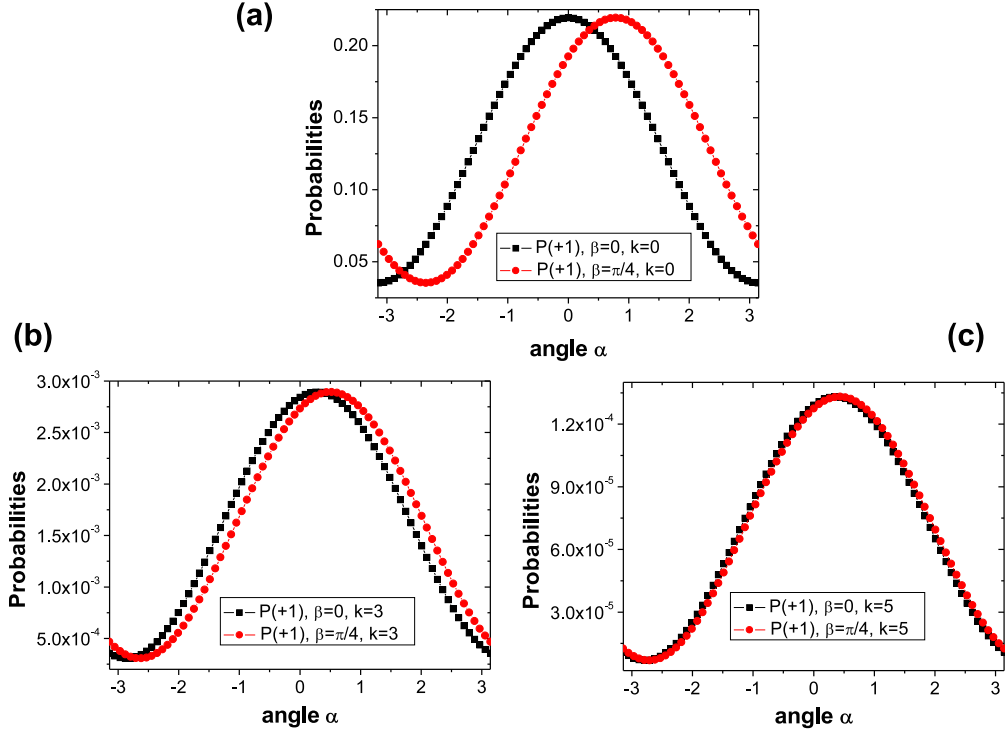}
\caption{(Color online) Fringe pattern as a function of the angle $\alpha$ of the polarization basis at the single-photon site.
The angle $\phi$ between the two bases of the pre-selection stage is set at $\phi = \pi/4$, while $g=1.2$. (a) Threshold $k=0$.
(b) Threshold $k=3$. (c) Threshold $k=5$. Square black points: fringe patterns
obtained by recording the $(+1)$ outcome at the measurement stage, where the measurement basis
$\{ \vec{\pi}_{\beta}, \vec{\pi}_{\beta_{\bot}} \}$ is set at $\beta = 0$. Circle red points: fringe patterns
obtained by recording the $(+1)$ outcome at the measurement stage, where the measurement basis
$\{ \vec{\pi}_{\beta}, \vec{\pi}_{\beta_{\bot}} \}$ is set at $\beta = \frac{\pi}{4}$. }
\label{fig:grafici_fringe_vs_k}
\end{figure*}
The obtained fringe patterns for the chosen case are reported in
Fig.\ref{fig:grafici_fringe_vs_k} and corresponds to the following
conditions. The $(+1)$ outcome of the dichotomic measurement is
recorded as a function of the polarization $\vec{\pi}_{\alpha}$ of
the injected single photon state. In particular, the two chosen
equatorial polarization bases $\{ \vec{\pi}_{\beta},
\vec{\pi}_{\beta_{\bot}} \}$ and $\{ \vec{\pi}_{\beta'},
\vec{\pi}_{\beta'_{\bot}} \}$ corresponds to $\beta = 0$ and
$\beta' = \frac{\pi}{4}$. We then analyzed three different choices
for the threshold $k$ at the pre-selection stage. When the
threshold $k$ is set to 0, the fringe pattern corresponding to the
two basis $\beta = 0$ and $\beta = \frac{\pi}{4}$ are mutually
shifted of an angle $\frac{\pi}{4}$, since no filtering and no
pre-selection is performed on the state. When the threshold $k$ is
increased, the mutual shift between the fringe pattern is
progressively reduced and cancelled, since a strong filtering of
the state is performed. In particular, the maximum of both the fringe
pattern in the $\beta = 0$ and $\beta = \frac{\pi}{4}$ bases is obtained for the $\vert
\Phi^{\overline{\alpha}} \rangle$ state with $\overline{\alpha} =
\frac{\pi}{8}$. This means that this pre-selection strategy for
sufficiently high value of $k$ enhances the correlations in the
micro-macro system in a specific polarization basis and suppresses
the correlations in the other bases. For this reason, the proposed
strategy does not allow to observe the violation of a Bell's
inequality in the micro-macro system here analyzed. The enhanced
value of the visibility could nevertheless be employed in quantum
lithography and quantum metrology schemes, in which high
visibility correlations pattern and high photon number regimes
are required. Recently it has indeed been shown how the
amplification process of a single photon probe can beat the
detrimental effect of losses which happen in the transmission and
detection stages \cite{Vite10b}. Such a scheme for non invasive
quantum metrology could benefits from the presented filtering
procedures in order to improve the visibility value of the
interference fringe pattern.

\section{Conclusions}

In this paper we have analyzed the properties of the macro states
obtained by a quantum injected amplification process, by
addressing the behavior of the distinguishability between
orthogonal macro-states when a filtering process is applied over a
portion of them. More specifically, we analyzed theoretically in
details several schemes for the realization of conditional
measurement-induced operations.
All these strategies are aimed at the manipulation and filtering
of the macro-states for their applications in different contexts, such
as the realization of a non-locality test or quantum communication.

We have identified a strategy, based on the ID device, able to
minimize the effects of the noise due to the vacuum injection into
the amplifier. The ID based filtering procedure is independent on
the analysis basis and selects the same portion of the state when
the measurement is performed in any equatorial polarization basis.

A different filtering procedure, based on the OF device, has been
deeply studied: it turned out that when a small portion of the
state is analyzed through the OF, the visibility of the overall
state, relative to a dichotomic measurement, is affected in a
different way depending on the polarization basis in which the
small portion has been measured. If the polarization basis is the
same of the macro qubit codification, the final visibility
increases with the increase of the filtering threshold, otherwise
it decreases. This behavior is related with the impossibility of
measuring non commuting operators on the same quantum state, as
explained in Section III.

We have further addressed in Sec.IV the trend of the macro state visibility
when an OF discrimination system is used even at the transmitted
state detection stage. In this case, the two OF apparatuses in both transmitted
and reflected branches play an opposing role in increasing or decreasing
the visibility of the fringe-pattern obtained in a micro-macro configuration.
Such analysis shows that the macro-states generated by optical parametric
amplification of a single-photon state are not suitable for quantum cryptography,
since they are not robust under an eavesdropping attack. Furthermore, we
showed that this discrimination method is not suitable for a micro-macro
non-locality test since it performs a base-dependent filtering of the detected
state.

Finally, in Sec.V we addressed a pre-selection scheme for the realization of
a Bell's inequality test which do not suffer the same detection loopholes
of the one based on post-selection strategies \cite{Vite10}. The proposed
method, based on the measurement of the reflected part of the wave-function
in two different bases, does not allow to violate a Bell's inequality,
since it induces the collapse of the correlations present in the
macro-states in only a single polarization basis.

Several open points remain to be investigated. The measurement-induced
operations analyzed in this paper are all based on dichotomic detection schemes.
Other approaches, such as the ones based on continuous variables measurements
or on the processes of coherent photon-addition and photon-subtraction, can lead to
a different manipulation of the QIOPA multiphoton states. Systems
with different properties from the one analyzed in this paper could be obtained
with these methods.

\section*{ACKNOWLEDGEMENTS}

We acknowledge support by the ``Futuro in Ricerca'' Project HYTEQ,
and Progetto d'Ateneo of Sapienza Universit\`{a} di Roma.


\begin{thebibliography}{10}

\bibitem{Fiur01}
J.~Fiurasek,
\newblock Phys. Rev. A {\bf 64}, 053817 (2001).

\bibitem{Heer06}
J.~Heersink {\em et~al.},
\newblock Phys. Rev. Lett. {\bf 96}, 253601 (2006).

\bibitem{Gloc06}
O.~Gl\"{o}ckl, U.~L. Andersen, R.~Filip, W.~P. Bowen, and G.~Leuchs,
\newblock Phys. Rev. Lett. {\bf 97}, 053601 (2006).

\bibitem{Ourj07a}
A.~Ourjoumtsev, H.~Jeong, R.~Tualle-Brouri, and P.~Grangier,
\newblock Nature {\bf 448}, 784 (2007).

\bibitem{Ourj09}
A.~Ourjoumtsev, F.~Ferreyrol, R.~Tualle-Brouri, and P.~Grangier,
\newblock Nature Physics {\bf 5}, 189 (2009).

\bibitem{Fili05}
R.~Filip, P.~Marek, and U.~L. Andersen,
\newblock Phys. Rev. A {\bf 71}, 042308 (2005).

\bibitem{Kita06}
A.~Kitagawa, M.~Takeoka, M.~Sasaki, and A.~Chefles,
\newblock Phys. Rev. A {\bf 73}, 042310 (2006).

\bibitem{Ourj07}
A.~Ourjoumtsev, A.~Dantan, R.~Tualle-Bruori, and P.~Grangier,
\newblock Phys. Rev. Lett. {\bf 98}, 030502 (2007).

\bibitem{Fiur09}
J.~Fiurasek,
\newblock Phys. Rev. A {\bf 80}, 053822 (2009).

\bibitem{Ferr10}
F.~Ferreyrol {\em et~al.},
\newblock Phys. Rev. Lett. {\bf 104}, 123603 (2010).

\bibitem{Xian10}
G.~Y. Xiang, T.~C. Ralph, A.~P. Lund, N.~Walk, and G.~J. Pryde,
\newblock Nature Photonics {\bf 4}, 316 (2010).

\bibitem{Zava10}
A.~Zavatta, J.~Fiurasek, and M.~Bellini,
\newblock Nature Phot., published online, doi:10.1038/nphoton.2010.260  (2010).

\bibitem{Usug10}
M.~A. Usuga {\em et~al.},
\newblock Nature Phys. {\bf 6}, 761  (2010).

\bibitem{DeMa05}
F.~{De Martini} and F.~Sciarrino,
\newblock Prog. Quant. Electr. {\bf 29}, 165 (2005).

\bibitem{Naga07}
E.~Nagali, T.~{De Angelis}, F.~Sciarrino, and F.~{De Martini},
\newblock Phys. Rev. A {\bf 76}, 042126 (2007).

\bibitem{DeMa08}
F.~{De Martini}, F.~Sciarrino, and C.~Vitelli,
\newblock Phys. Rev. Lett. {\bf 100}, 253601 (2008).

\bibitem{DeMa09}
F.~{De Martini}, F.~Sciarrino, and N.~Spagnolo,
\newblock Phys. Rev. Lett. {\bf 103}, 100501 (2009).

\bibitem{DeMa09a}
F.~{De Martini}, F.~Sciarrino, and N.~Spagnolo,
\newblock Phys. Rev. A {\bf 79}, 052305 (2009).

\bibitem{DeMa98}
F.~{De Martini},
\newblock Phys. Rev. Lett. {\bf 81}, 2842 (1998).

\bibitem{Vite10}
C.~Vitelli, N.~Spagnolo, L.~Toffoli, F.~Sciarrino, and F.~{De Martini},
\newblock Phys. Rev. A {\bf 81}, 032123 (2010).

\bibitem{Pope95}
S.~Popescu,
\newblock Phys. Rev. Lett. {\bf 74}, 2619 (1995).

\bibitem{Pawl09}
M.~Pawlowski, K.~Horodecki, P.~Horodecki, and R.~Horodecki,
\newblock arXiv:0902.2162  (2009).

\bibitem{Spag08}
N.~Spagnolo, C.~Vitelli, S.~Giacomini, F.~Sciarrino, and F.~{De Martini},
\newblock Opt. Express {\bf 16}, 17609 (2008).

\bibitem{DeMa09b}
F.~{De Martini},
\newblock Foundations of Physics,
\newblock doi:10.1007/s10701-010-9417-3, published online (2010), 
\newblock arXiv:0903.1992.

\bibitem{Stob09}
M.~Stobinska {\em et~al.},
\newblock arXiv:0909.1545v2  (2009).

\bibitem{Spag10}
N.~Spagnolo, C.~Vitelli, F.~Sciarrino, and F.~{De Martini},
\newblock Phys. Rev. A {\bf 82}, 052101 (2010).

\bibitem{Clau69}
J.~F. Clauser, M.~A. Horne, A.~Shimony, and R.~A. Holt,
\newblock Phys. Rev. Lett. {\bf 23}, 880 (1969).

\bibitem{Vite10b}
C.~Vitelli, N.~Spagnolo, L.~Toffoli, F.~Sciarrino, and F.~{De Martini},
\newblock Phys. Rev. Lett. {\bf 105}, 113602 (2010).

\end{thebibliography}
\end{document}